\documentclass[12pt,letter,useAMS]{aastex}
\usepackage{graphicx,amssymb}
\usepackage{natbib,txfonts}
\slugcomment{draft}

\bibpunct[, ]{(}{)}{;}{a}{}{,}
\def \mnras {MNRAS}
\def \pasp {PASP}
\def \apj {ApJ}
\def \apjs {ApJS}
\def \apjl {ApJL}
\def \aap {A\&A}
\def \nat {Nature}
\def \sci {Science}
\def \araa {ARAA}

\newcommand{\kms} {$\mathrm{ km \; s^{-1}}\,$}

\def\lesssim{\mathrel{\hbox{\rlap{\hbox{\lower4pt\hbox{$\sim$}}}\hbox{$<$}}}}
\def\gtrsim{\mathrel{\hbox{\rlap{\hbox{\lower4pt\hbox{$\sim$}}}\hbox{$>$}}}}

\long\def\symbolfootnote[#1]#2{\begingroup%
\def\thefootnote{\fnsymbol{footnote}}\footnote[#1]{#2}\endgroup} 

\begin{document}

%\cpright{}

%%%%%%%%%%%%%%%%%%%%%%%%%%%%%%%
%Title Page                   %
%%%%%%%%%%%%%%%%%%%%%%%%%%%%%%%
\title{The Early Asymmetries of Supernova~2008D / XRF 080109\footnote{Based on observations made with ESO Telescopes at the Paranal Observatory, under program 080.D-0107(A).}}
\author{Justyn R. Maund\altaffilmark{1,2,3}, J.~Craig~Wheeler\altaffilmark{4}, Dietrich Baade\altaffilmark{5}, Ferdinando Patat\altaffilmark{5},\\ Peter H\"oflich\altaffilmark{6}, Lifan Wang\altaffilmark{7} and Alejandro Clocchiatti\altaffilmark{8}}
\altaffiltext{1}{Dark Cosmology Centre, Niels Bohr Institute, University of Copenhagen, Juliane Maries Vej, DK-2100 Copenhagen \O, Denmark; justyn@dark-cosmology.dk}

\altaffiltext{2}{Department of Astronomy \& Astrophysics, University of California, Santa Cruz, 95064, U.S.A.}

\altaffiltext{3}{Sophie \& Tycho Brahe Fellow}

\altaffiltext{4}{Department of Astronomy and McDonald Observatory, The University of Texas, 1 University Station C1402, Austin, Texas 78712-0259, U.S.A.; wheel$@$astro.as.utexas.edu}

\altaffiltext{5}{ESO - European Organisation for Astronomical Research in the Southern Hemisphere, Karl-Schwarzschild-Str.\ 2, 85748 Garching b.\ M\"unchen, Germany; fpatat$@$eso.org; dbaade$@$eso.org}

\altaffiltext{6}{Department of Physics, Florida State University, Tallahassee, Florida 32306-4350, U.S.A.; pah$@$astro.physics.fsu.edu}

\altaffiltext{7}{Department of Physics, Texas A\&M University, College Station,
Texas 77843-4242, U.S.A.; wang$@$physics.tamu.edu}

\altaffiltext{8}{Departamento de Astronom$\mathrm{\acute{i}}$a y Astrof$\mathrm{\acute{i}}$sica, PUC Casilla 306, Santiago 22, Chile; aclocchi$@$astro.puc.cl}

%%%%%%%%%%%%%%%%%%%%%%%%%%%%%%%%
\begin{abstract}
  Spectropolarimetry of the Type Ib SN~2008D, associated with the
  XRF~080109, at two separate epochs, are presented.  The epochs of
  these observations correspond to V-band light curve maximum and 15
  days after light curve maximum (or 21 and 36 days after the XRF).
  We find SN~2008D to be significantly polarized, although the largest
  contribution is due to the interstellar polarization component of
  $Q_{ISP}=0\pm0.1\%$ and $U_{ISP}=-1.2\pm0.1\%$.  At the two epochs,
  the spectropolarimetry of SN~2008D is classified as being
  D1+L(\ion{He}{1})+L(\ion{Ca}{2}).  The intrinsic polarization of
  continuum wavelength regions is $<0.4\%$, at both epochs, implying
  an asymmetry of the photosphere of $<10\%$.  Similar to other Type
  Ibc SNe, such as 2005bf, 2006aj and 2007gr, we observed significant
  polarization corresponding to the spectral features of \ion{Ca}{2},
  \ion{He}{1}, \ion{Mg}{1}, \ion{Fe}{2} and, possibly, \ion{O}{1}\
  $\lambda 7774$, about a close-to-spherically-symmetric photosphere.
  We introduce a new plot showing the chemically distinct line forming
  regions in the ejecta and comment on the apparent ubiquity of highly
  polarized high-velocity \ion{Ca}{2} features in Type Ibc SNe.  The
  polarization angle of \ion{Ca}{2} IR triplet was significantly
  different, at both epochs, to those of the other species, suggesting
  high-velocity \ion{Ca}{2} forms in a separate part of the ejecta.
  The apparent structure in the outer layers of SN~2008D has
  implications for the interpretation of the early-time X-ray emission
  associated with shock break-out.  We present two scenarios, within
  the jet-torus paradigm, which explain the lack of an apparent
  geometry discontinuity between the two observations: 1) a jet which
  punched a hole straight through the progenitor and deposited Ni
  outside the ejecta and 2) a jet which stalled inside the radius of
  the photosphere as observed at the second epoch.  The lack of a
  peculiar polarization signature, suggesting strongly asymmetric
  excitation of the ejecta, and the reported properties of the
  shock-breakout favour the second scenario.
\end{abstract}

\section{Introduction}
\label{sec:intro}
\begin{figure}
\hfil
\rotatebox{-90}{
\includegraphics[width=7cm]{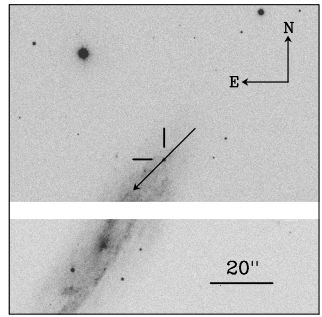}}
\hfil
\caption{The location of SN 2008D, indicated by the cross-hairs, in the 
galaxy NGC~2770.  The direction of the 
measured ISP component  (see \S\ref{sec:analysis:isp}) is shown by 
the arrow.  The image is 1 s in the V-band acquired with the VLT 
FORS1 and the gap in the image is due to a space between the two 
detector chips.} 
\label{fig:intro:posisp} \end{figure}

Core-collapse Supernovae (CCSNe) are inherently asymmetric events, due
to the asymmetric nature of the explosion mechanism \citep[see][for a
review]{2008ARA&A..46..433W}.  Probing the shapes of these events can
provide, therefore, important clues as to the nature of the explosion
mechanism and its influence on the standard photometric and
spectroscopic observables used to measure their properties.\\
Thomson scattering is a polarizing process, with the polarization of
the photon aligned in the direction orthogonal to the plane of
scattering (i.e. the plane defined by the ingoing and outgoing photon
directions and the point of scattering;  \citealt{Chandrasekhar}).  In SNe the principal site
for this process is at the photosphere, where the opacity is dominated
by free electrons.  For a resolved spherical
photosphere the most significant polarization would be seen to arise
from the limb, where only a narrow range of scattering angles
(for photons exiting the photosphere and being scattered into the line
of sight) are permissible.  For an unresolved spherical photosphere,
the polarized components arising from regions separated by $\pi / 2$
further around the photosphere are equal but opposite, cancelling out
and leading to a zero net observed polarization.  In the presence of
asymmetries, there is incomplete cancellation of the polarization
components arising from different points in the photosphere, leading
to a net observed polarization with the degree of polarization related
to the magnitude of the asymmetry and the polarization angle
corresponding to the orientation of the asymmetry on the sky (rotated
by $\pi / 2$).  This produces a wavelength-{\it independent} continuum
polarization.\\
The interaction of continuum photons with the line forming region also
produces a wavelength-{\it dependent} polarization.  Although resonant
scattering is inherently depolarizing, absorption along the line of
sight removes blue-shifted photons (leading to the standard absorption
component of P Cygni profiles).  If the line forming region does not
equally cover the photosphere, due to some large scale structure or
clumping, then unequal contributions of the polarization components
arising from the photosphere will be removed, leading to an increase
in polarization across the absorption and the so-called {\it inverted
  P-Cygni profiles} in the polarization spectrum
\citep{1988MNRAS.231..695C,1991ApJS...77..405J,2003ApJ...593..788K,2006NewAR..50..470H}.  The properties of the polarization at the absorption line are related
to the distribution of the line forming element about the photosphere.
Because photons scattered into the line of sight by resonant
scattering are depolarized, the red-shifted emission components are
inherently depolarized.  Polarimetry provides, therefore, a unique and
powerful way of directly measuring the shapes of SNe at early,
optically thick times.\\
Previous polarimetric studies of CCSNe have shown that the continua of
Type Ibc SN are generally more polarized at early times than Type II
SNe \citep{2001ApJ...550.1030W,2006Natur.440..505L, my2001ig,
  maund05bf, 2007A&A...475L...1M, 2008arXiv0806.1589T}.  In addition,
the increase in polarization across the absorption features of P~Cygni
profiles, coupled with the rotation of the polarization, have
demonstrated that the geometries of CCSNe are not described by just a
single global axial symmetry \citep{maund05bf}.\\
Here we report two epochs of VLT spectropolarimetry of the Type Ib SN,
associated with the X-ray Flash (XRF) 080109.  XRF 080109 was
discovered on 2008 Jan 10
\citep{2008GCN..7159....1B,2008Natur.453..469S}, in the galaxy
NGC~2770, a normal SA galaxy \citep{2008arXiv0807.0473T}. The location
of XRF 080109/SN~2008D, relative to its host galaxy, is shown as
Fig. \ref{fig:intro:posisp}.  The XRF was followed by an ultra-violet
transient.  An associated SN coincident with the XRF was identified
(2008D; \citealt{2008CBET.1202....1L}) and was spectroscopically
confirmed to be of Type Ib
\citep{2008Natur.453..469S,2008arXiv0805.1188M,2008arXiv0805.2201M}.
\citet{2008Natur.453..469S} determined that the XRF was
non-relativistic, unlike XRFs associated with the Gamma-Ray
Burst phenomenon, rather arising from the shock breakout arising from
a SN explosion.  The properties of the XRF suggested it arose from
breakout at a radius consistent with a Wolf-Rayet (WR) progenitor
star, into a dense wind ($\dot{M}=10^{-5}M_{\odot}$).
\citet{2008arXiv0807.1695M} and \citet{2008arXiv0807.1674T} modelled
the explosion as spherically symmetric, but both found that SN~2008D
was only slightly more energetic than normal Type Ibc SNe ($\sim 6
\times 10^{51}\mathrm{ergs}$) and less energetic than so-called
Hypernovae (e.g. SN~2002ap).  \citet{2008arXiv0807.1674T} found,
however, that poor fits of synthetic spectra to observed line profiles
indicated the role of significant asymmetries.  Here we directly establish those asymmetries by means of spectropolarimetry.\\
The paper is organised as follows: in Section \ref{sec:obs} we
describe our spectropolarimetric observations of SN~2008D.  The
results of these observations are presented in \S\ref{sec:results},
and these results are analysed in \S\ref{sec:analysis}.  The results
and analysis are discussed in \S\ref{sec:discussion} and we present
our conclusions in \S\ref{sec:conclusions}.

\setcounter{footnote}{0}
\section{Observations and Data Reduction}
\label{sec:obs}
\begin{figure}
\hfil
\rotatebox{-90}{
\includegraphics[width=14cm]{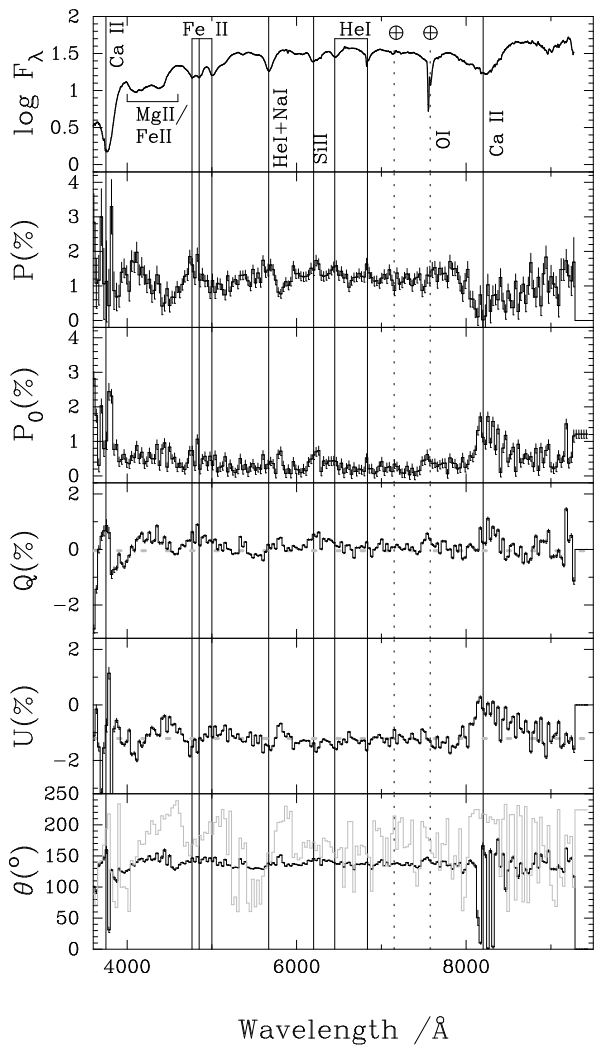}}
\hfil
\caption{Spectropolarimetry of SN~2008D at 2008 Jan 31.22.  The six
  panels (from top to bottom) give: the observed flux spectrum, in
  logarithmic flux units ($\mathrm{ergs\,s^{-1}\,cm^{-2}\,\AA^{-1}}$),
  uncorrected for reddening; the observed polarization spectrum ($P$);
  the observed polarization spectrum corrected for the ISP ($P_{0}$;
  see \S\ref{sec:analysis:isp}); the Stokes $Q$ and $U$ parameters
  uncorrected for the ISP (the location of the ISP is shown for each
  panel by the dashed grey line); and the polarization angle $\theta$,
  uncorrected for the ISP in black and corrected in grey.  Line identifications (at
  absorption minimum) are provided in the top panel, and are based on
  the identifications of \citet{2008arXiv0805.1188M} and \citet{2008arXiv0805.2201M}.  The data are
  corrected for the recessional velocity of the host galaxy and have
  been rebinned to 30$\mathrm{\AA}$\ for clarity.}
\label{fig:obs:epoch1}
\end{figure}

\begin{figure}

\hfil
\rotatebox{-90}{
\includegraphics[width=14cm]{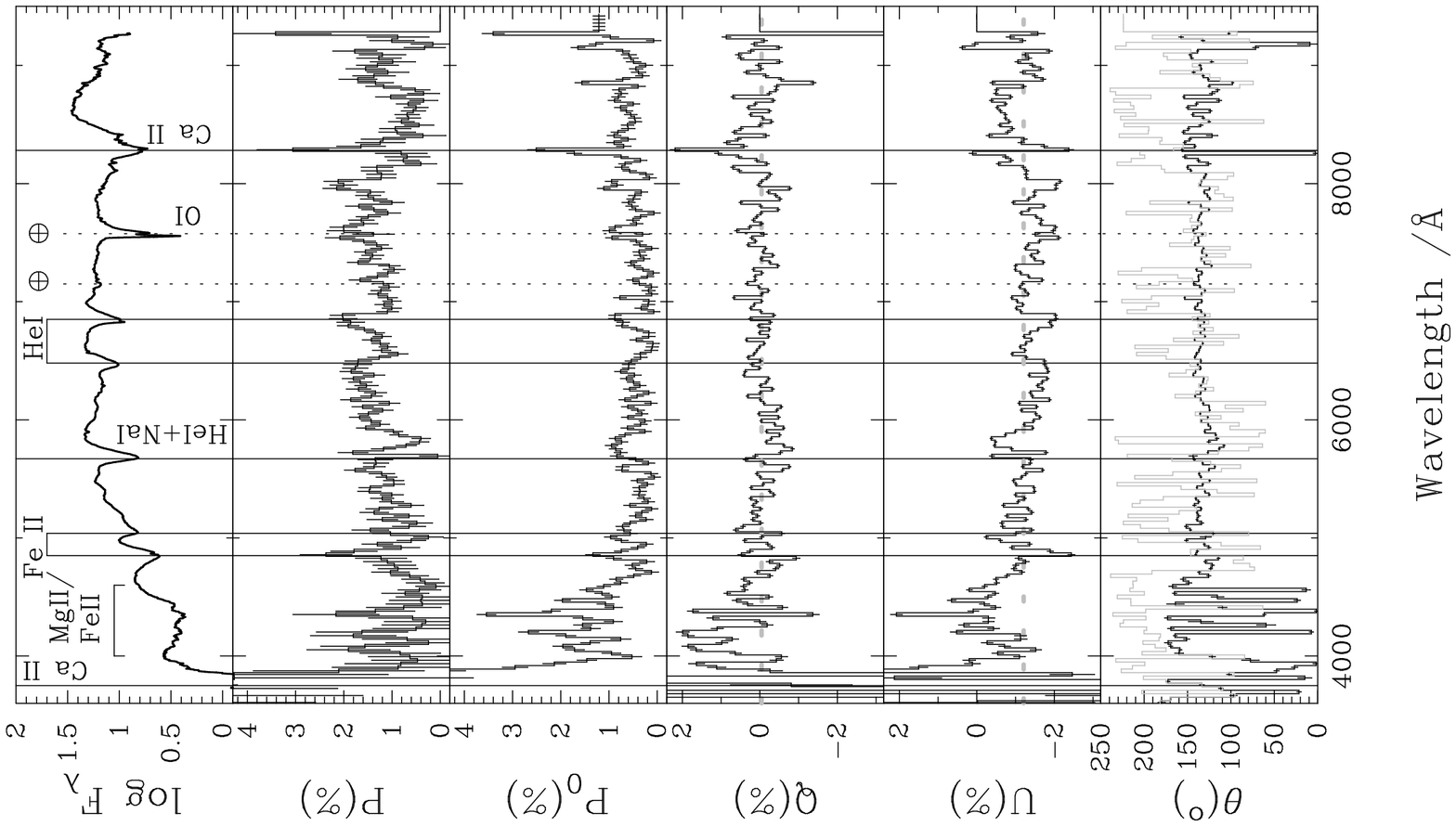}}
\hfil
\caption{The same as for Fig. \ref{fig:obs:epoch1} but for spectropolarimetry of SN~2008D at 2008 Feb 15.18.}
\label{fig:obs:epoch2}
\end{figure}

Spectropolarimetry of SN~2008D was conducted at two epochs, 2008 Jan
31 and 2008 Feb 15 (corresponding to approximately V-band light-curve
maximum and 15 days after maximum, respectively;
\citealt{2008arXiv0805.1188M}), using the ESO VLT FORS1 instrument in
the PMOS mode \citep{1998Msngr..94....1A}.  A journal of these
observations is shown as Table \ref{tab:obs:log}.  All observations
were conducted with the 300V grism, providing a wavelength range of
$\mathrm{3650-9300\AA}$, and with no order separation filter used.
Observations of SN~2008D, at both epochs, consisted of two sets of
four exposures at each of the four half-wavelength retarder plate
angles.  In addition, at each epoch, a flux standard was observed with
the full polarimetry optics in place at a single retarder plate angle
($0\degr$).  The observations were reduced using {\sc
  iraf}\footnote{http://iraf.noao.edu - IRAF is distributed by the
  National Optical Astronomy Observatories, which are operated by the
  Association of Universities for Research in Astronomy, Inc., under
  cooperative agreement with the National Science Foundation.} and
our own specially written software, following the scheme presented by
\citet{maund05bf}.\\
The spectropolarimetric observations at each epoch were combined using
a weighted averaging scheme, with the Stokes parameters weighted by
the flux.  Due to its relatively high declination, SN~2008D was only
observed at low altitude ($30\degr$) by the VLT.  The FORS1 instrument
uses a mosaic of two thinned, backside-illuminated 2k$\times$4k E2V
CCD44-82 detectors, with increased sensitivity in the blue, but severe
fringing and second-order contamination at $\mathrm{>6500\AA}$
(c.f. spectropolarimetry with the previous Tektronix CCD detector;
e.g. \citealt{my2001ig}).  The nature of the observations, of a highly
reddened object at high-airmass with the slit oriented at $PA=0\degr$
rather than the parallactic angle, was beneficial in that it
significantly reduced the flux in the blue (see
Figs. \ref{fig:obs:epoch1} and \ref{fig:obs:epoch2}) and, hence,
minimised order contamination of blue flux in the red (and hence
contamination of the resulting measured polarization).  Since polarimetry
uses a relative measurement of flux, the severe reddening of the
spectrum has no effect on the final measured degree of polarization.\\
In order to assess the role of any instrumental changes, including
evolution of the fringing pattern, the instrumental signature
corrections ($\epsilon_{Q}$ and $\epsilon_{U}$;
\citealt{2008A&A...481..913M}) were compared for each pair of datasets
and found to be consistent to within the measurement error of $\ll
0.1\%$.\\
The spectra were flux calibrated, with respect to the observed flux
standard, but were {\it not} corrected for atmospheric extinction
(such that the flux spectra presented here are significantly redder
than presented elsewhere).  The calibration of the PMOS mode of FORS1
was checked using observations of the polarized
standard star Vela~1~95.  The data were corrected for the heliocentric recessional velocity of the host galaxy of $1\,947$\ \kms \footnote{NED: http://nedwww.ipac.caltech.edu}.\\
Synthetic broad V-band polarimetry was measured from the data,
weighting the observed spectropolarimetry as a function of wavelength
with the Johnson V passband response function.

\section{Observational Results}
\label{sec:results}
The observed spectropolarimetry of SN~2008D at 31 Jan 2008 and 15 Feb
2008 is presented as Figs. \ref{fig:obs:epoch1} and
\ref{fig:obs:epoch2}.

\subsection{General Spectroscopic Properties}
\label{sec:results:spec}
We summarise the spectroscopic properties of SN~2008D at the two
epochs of our observations, to serve as an orientation for the
discussion of the polarimetric properties of this object.  Previously,
spectra of SN~2008D have been presented by
\citet{2008Natur.453..469S}, \citet{2008arXiv0805.1188M},
\citet{2008arXiv0805.2201M}, \citet{2008arXiv0807.1695M} and
\citet{2008arXiv0807.1674T}.  Here we adopt the line identifications
made by \citet{2008arXiv0805.1188M} and
\citet{2008arXiv0805.2201M}.\\
At the two epochs of our observations the spectrum of SN~2008D is
composed of broad P~Cygni profiles of \ion{He}{1}, \ion{Ca}{2},
\ion{O}{1} and \ion{Fe}{2}.\\
\ion{Ca}{2} H\&K and the IR triplet (IR3) are present at the blue and
red extremes of the spectra.  At the first epoch, the full P~Cygni
profile of \ion{Ca}{2} H\&K is observed, with the absorption minimum
at -14\,100\,\kms.  In addition, weak narrow absorptions are also
observed superposed on the emission feature, at the rest wavelength in
the host galaxy, arising in the Interstellar Medium.  At the second
epoch, the absorption component is no longer detected, due to the
spectrum becoming redder and steeper, such that only the emission
component and the decline towards the minimum are observed.  The
\ion{Ca}{2} IR3 feature is detected at both epochs.  At the first
epoch, the emission component is truncated in the red by an \ion{O}{1}
line \citep{2008arXiv0805.2201M}.  The absorption component,
corresponding to a velocity of $-12\,000$ \kms, is a broad
single-minimum feature.  At the second epoch there is little evidence
of the redward absorption due to \ion{O}{1}, but the absorption
minimum of \ion{Ca}{2} IR3 is separated into seemingly high-velocity
(HV) and low-velocity minima at $-10\,200$ \kms and $-6\,400$ \kms,
respectively (but note that the HV component is moving at lower
velocities than the
single component observed at the first epoch).\\
Similar to the spectrum of SN~2005bf, \ion{Fe}{2} (42) is observed as
separate resolved absorption minima at the first epoch ($v \sim
9\,500$ \kms); at the second epoch, due to a general decrease in
observed velocities the \ion{Fe}{2} features are blended into a single
profile, with a likely significant contribution from \ion{He}{1}
$\lambda 5015$, commensurate with the observed increase in the
strength of other \ion{He}{1} lines
elsewhere in the spectrum.  A combined \ion{Mg}{2} $\lambda 4471$ and \ion{Fe}{2}(37,38) feature is observed at $\sim 4400 \AA$.\\
At the first epoch a set of \ion{He}{1} lines ($\lambda \lambda 5876,
6678, 7065$) are observed in the spectra, with the two redder lines
being observed as slight notches (corresponding to $v_{He} = -10 \,
100$ \kms).  The stronger \ion{He}{1} $\lambda 5876$ line may also be
blended with \ion{Na}{1}, which may explain the apparent disparity
between line strengths in the \ion{He}{1} series.  At the second epoch
the \ion{He}{1} lines have significantly increased in strength,
although with a slight decrease in velocity to $-9\,500$ \kms.  A
narrow absorption is observed at the rest
wavelength of \ion{Na}{1}~D at the host galaxy. \\
At the first epoch, \ion{Si}{2} $\lambda 6355$ is observed as a
shallow P~Cygni profile, that is potentially blended with a
high-velocity component of $\mathrm{H\alpha}$\,
\citep{2008arXiv0805.1188M}.  Assuming this feature is purely due to
\ion{Si}{2} the absorption minimum corresponds to $-7\,790$ \kms.
This feature is absent from the data of the second epoch.  There is a
hint of \ion{O}{1} $\lambda 7774$ in emission, with the corresponding
absorption trough coincident with the nearby telluric feature
(suggesting an approximate velocity $\sim -8 \, 000$ \kms at the first
epoch, and $v \sim 7\,600$ \kms at the second).
\citet{2008arXiv0805.2201M} observe the full P~Cygni profile of the
\ion{O}{1} line in their spectroscopic data, at 2 days prior to our
first
observation.\\
The Galactic foreground reddening to SN~2008D is given as
$E(B-V)_{MW}=0.02 3$ \citep{schleg98}.  The equivalent width of
\ion{Na}{1}~D measured from our low resolution data is a poor
indicator of reddening arising in the host, as previous
high-resolution studies of SN~2008D have shown it to be a
superposition of a number of different, saturated absorption systems
\citep{2008Natur.453..469S,2008arXiv0805.2201M}.  Previous studies
have determined the reddening to fall in the range $E(B-V)=0.6-0.9$
\citep{2008Natur.453..469S,2008arXiv0805.1188M,2008arXiv0805.2201M,2008arXiv0807.1695M,2008arXiv0807.0473T}.
Here, we adopt the value $E(B-V)=0.65$, the most commonly used value,
and assume the foreground reddening is negligible.
\subsection{General Spectropolarimetric Properties} 
\label{sec:results:pol}

\begin{figure}
\hfil
\rotatebox{-90}{
\includegraphics[width=7cm]{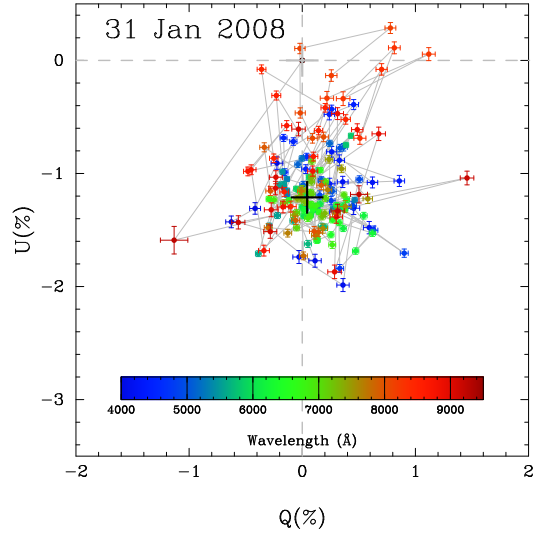}}
\hfil
\caption{Spectropolarimetry of SN~2008D at 2008 Jan 31.22 on the Stokes $Q-U$ plane, uncorrected for the ISP.  The location of the ISP is indicated by the black cross, and the zero polarization is indicated by the grey cross and dashed lines.  The observed data are colour coded according to wavelength, following the scheme in the colour bar. }
\label{fig:obs:epoch1qu}
\end{figure}

\begin{figure}
\hfil
\rotatebox{-90}{
\includegraphics[width=7cm]{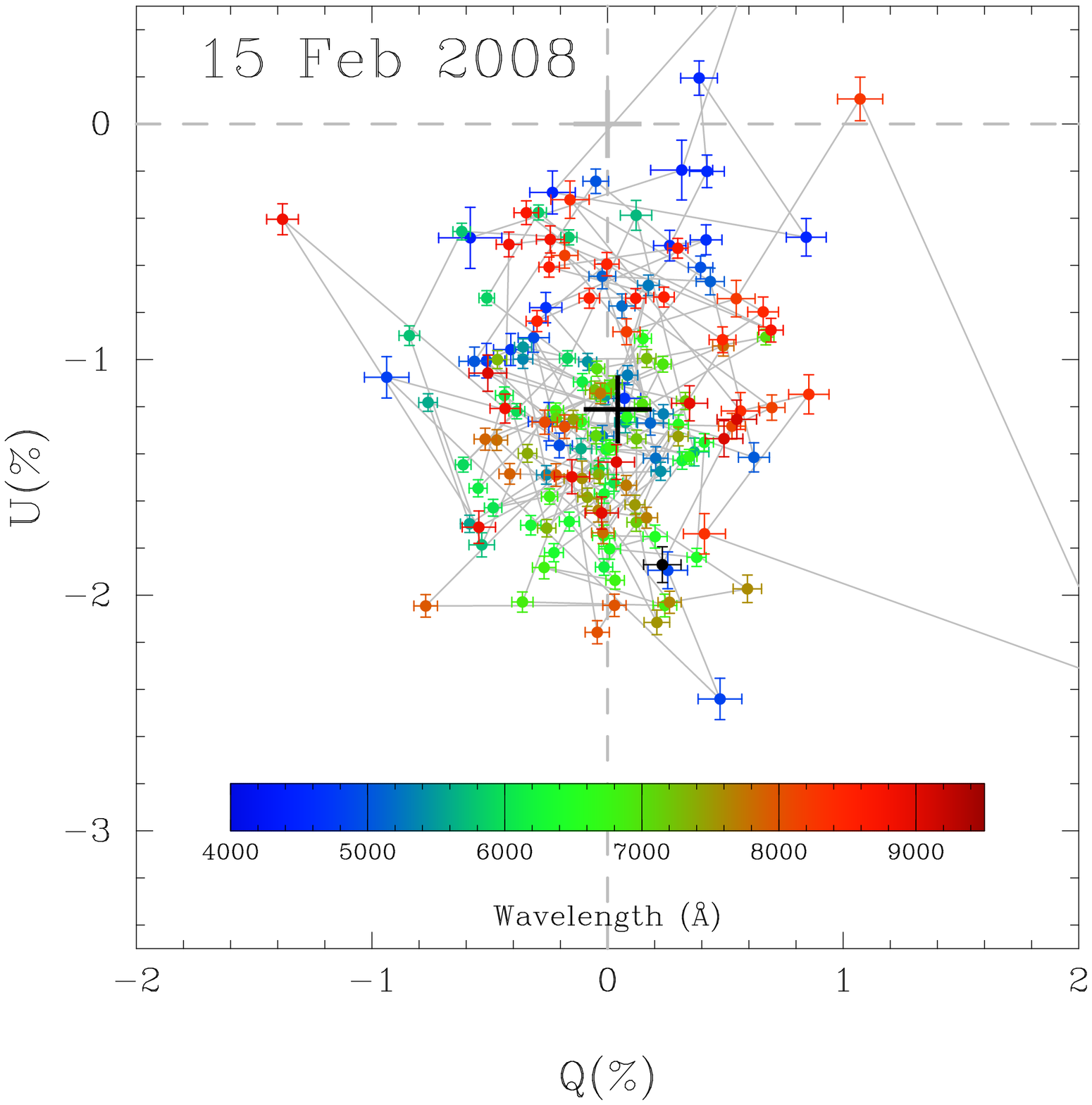}}
\hfil
\caption{The same as for Fig. \ref{fig:obs:epoch1qu}, but for spectropolarimetry of SN~2008D at 2008 Feb 15.18.}
\label{fig:obs:epoch2qu}
\end{figure}

The spectropolarimetric data are plotted on the Stokes $Q-U$ plane for
both epochs on Figs. \ref{fig:obs:epoch1qu} and
\ref{fig:obs:epoch2qu}.  SN~2008D is observed to be significantly
polarized at both epochs, with a baseline polarization of $P \sim 1.3\%$.
As shown on Figs. \ref{fig:obs:epoch1} and \ref{fig:obs:epoch2} this
level of polarization is common at both epochs.  Most of this
polarization is in the $U$ Stokes parameter, as evident in
Figs. \ref{fig:obs:epoch1qu} and \ref{fig:obs:epoch2qu}, as the data
points are clustered in the same location of the Stokes $Q-U$ plane
and the polarization angle is $\sim 135 \degr$ across most of the
spectrum.  The apparent rise in the polarization at the blue and red
extremes of the data is consistent with an increase in the
uncertainties on the measured polarization, due to declining
signal-to-noise in these portions of the spectrum and the possible
wavelength dependence of the the Interstellar Polarization (ISP).\\
The most significant deviations from this baseline level of
polarization are associated with spectral features.  At the first
epoch a peak polarization of $3\pm0.8\%$ is observed at the
\ion{Ca}{2} H\&K absorption component.  This is contrasted for the
\ion{Ca}{2} IR3 absorption, for which the absorption minimum is
associated with a minimum (0.6\%) in the polarization spectrum.  On
the $Q-U$ plane, the data corresponding to the \ion{Ca}{2} IR3
absorption is observed to lie distinctly separate from the rest of the
data.  Significant polarization is associated with the \ion{He}{1}
$\lambda 5876$ line, and the redder He lines are also polarized albeit
at a lower degree.  With the increase in the strength of the
\ion{He}{1} lines at the second epoch, there is a correlated increase
in the polarization.  The resolved absorption components of the
\ion{Fe}{2}(42) multiplet and the \ion{Mg}{2} $\lambda 4471$ line, at
the first epoch, are associated with strong polarization at the
level of $1.8\pm0.2\%$.\\
Given the substantial degree of reddening associated with this SN (see
\S\ref{sec:results:spec}), there is the expectation that the ISP
component is non-zero.  As has been demonstrated in a number of cases,
the determination and subtraction of the ISP component is vital to
determining the intrinsic polarization ($P_{0}$) and, hence, geometry
of SNe.  Importantly, the ISP can cause intrinsically unpolarized
features to be observed to be polarized and make polarized features
appear depolarized.  The standard expectation for P~Cygni profiles, in
the flux spectrum, is that significant polarization is associated with
the absorption component while the emission component is
unpolarized. We see that, at the first epoch, the absorption component
of \ion{Ca}{2} IR3 is unpolarized.  Furthermore, we note that the
general behaviour of the data on the Stokes $Q-U$ plane is constant with time,
occupying the same location (despite variability in the polarization
associated with spectral lines).\\
Synthetic V-band polarimetry of this data yielded $p_{V}=1.2\pm0.1\%$
($\theta=135\degr$) and $1.1\pm0.1\%$ ($131\degr$), at 31 Jan 2008 and
15 Feb 2008 respectively.  \citet{2008arXiv0810.4333G} present broad V-band
polarimetric measurements which, not only agreeing with our synthetic
V-band polarimetry, also show that SN~2008D occupies the same location
of the $Q-U$ plane prior to and following our two observations (see
\S\ref{sec:discussion}).
\section{Analysis}
\label{sec:analysis}
\subsection{Interstellar Polarization}
\label{sec:analysis:isp}
The determination of the ISP is crucial to determining the intrinsic
polarization of the SN.  The reddening towards SN~2008D can be used to
provide a constraint on the maximum degree of ISP. The total reddening
to SN~2008D (see \S\ref{sec:results:spec}) constrains the ISP to be
$p_{ISP}<5.85\%$ (assuming a standard Serkowski-Galactic type ISP),
arising predominantly in the host galaxy.  In order to make a precise
estimate of the ISP component particular assumptions must be used,
principally based on there being portions of the spectrum that are
intrinsically unpolarized, such that the observed polarization of
these regions is due to the ISP alone. We attempt to remove only a
single total ISP, rather than account for individual ISPs arising in
the Galaxy and the host galaxy, as the total ISP is a vector sum of
all individual components.  A number of techniques, based on this
primary assumption, have been used and we discuss in
turn the alternative values of the ISP indicated.\\
Under the assumption that particular wavelength regions of the
spectrum are intrinsically depolarized due to the blending of numerous
overlapping Fe lines in the range 4800-5600\AA,
\citet{2001ApJ...556..302H} used the observed polarization properties
at particular wavelengths as measures of the ISP. The \ion{Fe}{2}
multiplet 42 lines are clearly resolved at both epochs
\citep{maund05bf}, however immediately redward in the range
5200-5500\AA\ there are a number of overlapping features.  Under the
assumption that the line blanketing opacity dominates over electron
scattering in this wavelength range, such that this region is
intrinsically depolarized, we measure the median average values of the
Stokes parameters as $Q_{ISP}=-0.085 \pm 0.234\%$ and $U_{ISP}=-1.164
\pm 0.175 \%$ at the first epoch and $Q_{ISP}=-0.061 \pm 0.240\%$ and
$U_{ISP}=-1.164 \pm 0.266 \%$ at the second epoch.  These values of
the ISP Stokes parameters are identical, within the uncertainties, and the average value corresponds to $p_{ISP}=1.16\pm0.15\%$ and $\theta_{ISP}=133\pm4\degr$.\\
An additional measurement can be derived using the technique of
\citet{1997PASP..109..489T}, under the assumption that the emission
component of a P~Cygni profile is intrinsically unpolarized.  The
polarization of the emission component, corrected for the polarization
of the continuum (actually a mixture of the intrinsic continuum
polarization and the ISP), reflects the ISP component alone and can be
measured by a least-squares fit to the equations $Q_{tot}\times
F_{tot}=Q_{cont}\times F_{cont}+Q_{line}\times F_{line}$ and
$U_{tot}\times F_{tot}=U_{cont}\times F_{cont}+U_{line}\times
F_{line}$ to determine the Stokes parameters of the emission line
$Q_{line}$ and $U_{line}$ (in the presence of no ISP
$Q_{line}=U_{line}=0$).  This technique was applied to the emission
components of the \ion{Mg}{1} $\lambda$4470 / \ion{Fe}{2} and
\ion{Ca}{2} IR3 lines at the second epoch, where a good approximation
for the continuum level could be made.  At the first epoch, the
emission components of these lines are truncated by blends with other
features, whereas at the second epoch these features appear unblended.
The ISP component was measured for both lines to be
$Q_{ISP}=0\pm0.1\%$ and $U_{ISP}=-1.3\pm0.1\%$ (where the quoted error
is the scatter in fits using different wavelength ranges and
different continuum levels).\\
The fact that the ISP is time-invariant supports these values, given
the apparent constant level of polarization and the constant location
of the data on the $Q-U$ plane at the two observational epochs.
\citet{2008arXiv0810.4333G} observed similar Stokes parameters with
their broad-band polarimetry, between 3.6 and 78.5 days
post-explosion, suggestive that the bulk of the observed polarization
(excluding polarization associated with strong
lines) was due to a unvarying ISP.\\
In general, the ISP is expected to be wavelength-dependent, but
characterising this dependence is difficult for objects such as
SN~2008D which show strong intrinsic wavelength-dependent polarization
associated with spectral lines.  For a Serkowski-type ISP, for Milky
Way-like dust, the difference in ISP between the peak polarization, at
a characteristic wavelength of $5500\AA$, and the ISP polarization
measured for the emission components of the \ion{Ca}{2} lines at the
blue and red ends of the spectrum is smaller than the error of the
measured Stokes parameters of the ISP.  The small degree of wavelength
dependence justifies the use of a single-valued, wavelength
independent ISP as inferred from this data.  For a number of SNe, the
wavelength-dependence of the ISP, arising from their host galaxies,
has been observed to deviate significantly from a standard
\citeauthor{1975ApJ...196..261S}-type ISP law
\citep{2001PASP..113..920L,maund05bf,nando2006X}.  In the case of
SN~2008D, there is no evidence for the magnitude of the polarization
rising significantly at either the blue or red extremes of the
observed spectrum (suggesting the peak in the ISP occurs within the
wavelength range of these observations).  In fact, the baseline
polarization of the data is approximately constant across the observed
wavelength range, with a
possible peak around $5500\AA$ (see above), which lends further credence to the to the assumption that the ISP can be considered approximately wavelength independent for our observations (with any wavelength dependence of the ISP over the wavelength range being smaller than the uncertainties of our wavelength independent estimate).\\
The weighted average of the measured Stokes parameters for the ISP,
from the different methods, are $Q_{ISP}=-0.04\pm0.13\%$ and
$U_{ISP}=-1.21\pm0.12$, which correspond to $p_{ISP}=1.21\pm0.12\%$
and $\theta_{ISP}=134\pm4.3\degr$.  The direction of the ISP vector is
shown on Fig. \ref{fig:intro:posisp}.  Contrary to the standard
prediction that the ISP vector be aligned with the spiral arms of the
host galaxy, due to the alignment of dust grains by magnetic field
lines parallel to the spiral arm \citep{1987MNRAS.224..299S}, the ISP
is not aligned with the nearest spiral arm.  This may indicate that
the ISP determined may either be incorrect, that the magnetic field in
the host is not aligned with the spiral arms at large radii or that
the ISP is the sum of a number of separate components, perhaps arising
in the Galaxy, the host galaxy, the intervening
medium and, importantly, in the circumstellar environment of the SN.\\

\subsection{The intrinsic polarization of SN~2008D}
\label{sec:analysis:intrinsic}

\subsubsection{The continuum}
\label{sec:analysis:intrinsic:cont}
After correcting the observed data for the ISP, we note that away from
obvious spectral features the intrinsic polarization is generally
$<0.4\%$.  Furthermore, at both epochs the wavelength region
$7100-7500\AA$\ appears to have a relatively constant low
polarization, with only weak features appearing in the flux
spectrum. Using this region as representative of the continuum, we measure the intrinsic polarization to be $0.22 \pm 0.14\%$ and
$0.21 \pm 0.17\%$ at the first and second epochs respectively (where
these values are weighted means over the data in this wavelength
range).  Assuming a spheroidal photosphere, this polarization
corresponds to an apparent axial ratio $>0.9$
\citep{1991A&A...246..481H}.

\subsubsection{\ion{He}{1} $\lambda\lambda 5876,6678,7065$}
\label{sec:analysis:intrinsic:hei}
\begin{figure}
\hfil
\rotatebox{-90}{
\includegraphics[width=7cm]{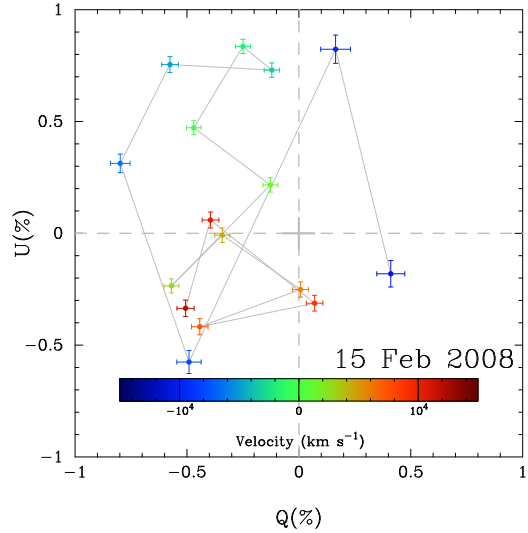}}
\hfil
\caption{The polarization across the \ion{He}{1} $\lambda 5876$ line
  at 2008 Feb 15.  The points are colour coded according to velocity
  with respect to the rest wavelength of the line.  The data have been
  corrected for the recessional velocity of the host galaxy and the ISP.
  The data marked in blue correspond to the polarized absorption
  component of the line profile.}
\label{fig:obs:heline}
\end{figure}

Significant intrinsic polarization is associated with the \ion{He}{1} lines at both epochs.  At the second epoch, higher levels of
polarization are measured for the redder \ion{He}{1} lines, in line
with the absorption becoming stronger (i.e. the fraction of continuum
photons being removed by the line is larger at the second epoch).  If
the line forming region, for a general line,1 blocks all the unpolarized light from a spheroidal photosphere, then the maximum degree of
polarization expected at absorption minimum ($p_{trough}$) is:
\begin{equation}
p_{trough} \leq p_{cont} \cdot \frac{F_{cont}}{F_{trough}} 
\label{eq:res:leonard}
\end{equation}
following the prescription of \citet{2002PASP..114.1333L}, where
$p_{cont}$ is the degree of polarization measured in adjacent
``continuum'' regions and $F_{cont}$ and $F_{trough}$ are the fluxes
measured at the nearby continuum and at the absorption component
minimum, respectively.  There is some difficulty with the measurement
of the maximum polarization of \ion{He}{1} $\lambda 5876$ in the
intrinsic polarization spectrum, as there is variability at and around
the position corresponding to the absorption line minimum in the flux
spectrum.  We note, however, that the redder \ion{He}{1} lines show
the expected inverted P-Cygni profile in the polarization spectrum.
At the second epoch, for $\lambda\lambda 5876,6678,7065$ we estimate theoretical
polarization limits of $0.68\%$, $0.44\%$ and $0.45\%$, respectively,
as opposed to the observed values of $1.11\%$, $0.58\%$, and $0.73\%$.
\citet{2001ApJ...553..861L}, \citet{2002PASP..114.1333L} and
\citet{2008arXiv0806.1589T} present a number of reasons why this
theoretical limit may not be actually observed: depolarization of
resonantly scattered continuum photons in optically thick lines, the
contribution of strong emission line
components or clumpy ejecta.\\
A significant clue as to the true reason
behind the significant polarization is observed in the rotation of the
polarization angle across the \ion{He}{1} $\lambda 5876$ line (and to
a lesser extent for $\lambda\lambda 6678,7065$).  This rotation
corresponds to a velocity-dependent loop structure, which is shown as
Fig. \ref{fig:obs:heline}.  Such a loop feature is generally
interpreted as a departure from a single global axial symmetry, such
that the axis of symmetry of the line forming region is not aligned
with that of the photosphere and the line forming region also only
covers a fraction of the photosphere
\citep{2003ApJ...593..788K,maund05bf}.  The similarity of the
polarization angle of three \ion{He}{1} lines,
$\theta_{5876}=141\pm9\degr$, $\theta_{6678}=147\pm9\degr$ and
$\theta_{7065}=152\pm7\degr$, at the first epoch suggest a common
geometry for the line forming regions of these lines.  The two red
\ion{He}{1} lines have similar polarization angles at the second epoch,
$\theta_{6678}=147\pm7\degr$ and
$\theta_{7065}=154\pm5\degr$, suggesting the \ion{He}{1} line forming region is the same as in the first epoch.  The polarization angle associated with \ion{He}{1} $\lambda 5876$ changes to $39\pm7\degr$, suggesting possible blending due to \ion{Na}{1} D.\\
At the second epoch, \ion{He}{1} $\lambda 5015$ is also expected to be
a prominent feature, although coincident with a number of \ion{Fe}{2}
lines.  The presence of this line is suggested by the similar
polarization angle ($147\pm6\degr$) associated with the absorption at 4890\AA.\\
We conclude that the violation of Eq. \ref{eq:res:leonard} and the observed rotation of the polarization angle for the \ion{He}{1} lines is due to the lines arising in an excited \ion{He}{1} clump that is off-axis relative to the axis of symmetry of the photosphere and does not cover the full area of the photosphere.

\subsubsection{\ion{Ca}{2}}
\label{sec:analysis:intrinsic:caii}
\begin{figure}
\hfil
\rotatebox{-90}{
\includegraphics[width=7cm]{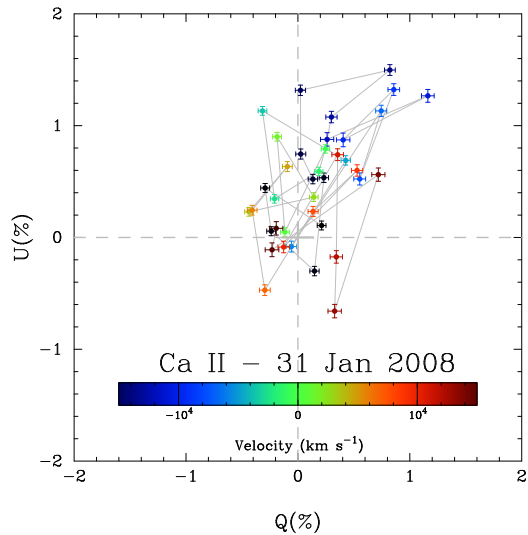}}
\hfil
\caption{The same as Fig. \ref{fig:obs:heline} but for the \ion{Ca}{2} IR3 feature at 2008 Jan 31.}
\label{fig:obs:caloop1}
\end{figure}

\begin{figure}
\hfil
\rotatebox{-90}{
\includegraphics[width=7cm]{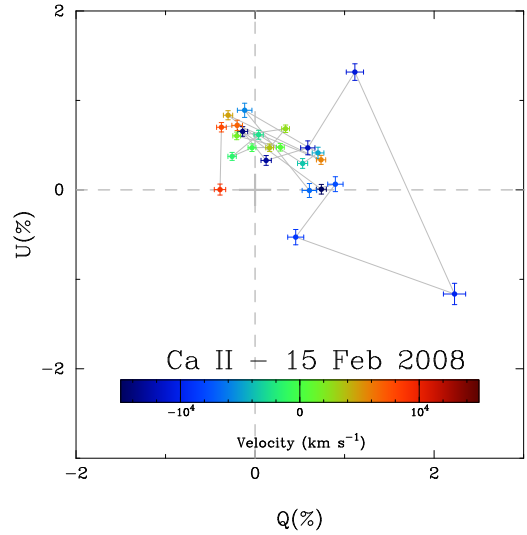}}
\hfil
\caption{The same as Fig. \ref{fig:obs:heline} but for the \ion{Ca}{2} IR3 feature at 2008 Feb 15.}
\label{fig:obs:caloop2}
\end{figure}

At the first epoch, \ion{Ca}{2} H\&K is observed to be polarized (with
an inverted P-Cygni profile) at $2.4\pm0.8\%$, a similar level to that
observed for the \ion{Ca}{2} IR3 at $1.8\pm0.3\%$.  The \ion{Ca}{2}
IR3 absorption feature is observed to be polarized at both epochs,
with the subtraction of the ISP, producing the expected ``inverted
P-Cygni profiles'' in the polarization spectra.  As it is unclear if
the flux at the absorption minimum of the \ion{Ca}{2} H\&K feature is
significantly detected at the second epoch (\S\ref{sec:results:spec}),
we cannot ascertain the real degree or significance of the observed
polarization around the wavelength of that feature at that epoch.
At the second epoch, in which a probable HV component is observed, a
polarization of \ion{Ca}{2} IR3 is observed to be $2.5\pm0.7\%$.  This
suggests the apparent HV component at the second epoch also produces
the absorption minimum at the first epoch, and the two components were
not resolved at the first epoch.  Similarly to
\S\ref{sec:analysis:intrinsic:hei}, the theoretical limiting
polarization for \ion{Ca}{2} is $0.92\%$, such that the observed data
suggests a similar scenario for \ion{Ca}{2} (as suggested for
\ion{He}{1}) being formed in a region with a different axial symmetry
to that of the photosphere and incomplete coverage.  At the first epoch the polarization angle
at the \ion{Ca}{2} IR3 absorption minimum is $37\pm4\degr$, whereas at
the second epoch the angle is $25\pm3\degr$.  Importantly, these angles
are significantly different than those observed for \ion{He}{1},
\ion{O}{1} and \ion{Fe}{2}, suggesting the \ion{Ca}{2} IR3 line forms
in a significantly different portion of
the ejecta to the lines of the other species.\\
At the first epoch, the absorption component of IR3 is obviously
polarized, although the data is not of sufficient quality (due to the
severity of fringes) to discern a loop (see
Fig. \ref{fig:obs:caloop1}).  By the time of the second epoch, a loop
is easily identified (Fig. \ref{fig:obs:caloop2}) corresponding to the
HV component of the absorption profile.

\subsubsection{\ion{Fe}{2}}
\label{sec:analysis:intrinsic:feii}
The observed level of polarization associated with the \ion{Fe}{2}(42)
lines is significantly reduced upon subtraction of the ISP, as
indicated in Figs. \ref{fig:obs:epoch1} and \ref{fig:obs:epoch2}.
Some polarization is still associated with these lines at the first
epoch ($0.6 - 1 \pm 0.2\%$), but a larger polarization ($1\pm0.3\%$)
is observed at this wavelength at the second epoch, due to the increase in strength of the \ion{He}{1} $\lambda 5015$ line (see \S\ref{sec:analysis:intrinsic:hei}).\\
At first epoch, the peak polarization for the bluest absorption of
\ion{Fe}{2}(37,38), at $-9\,600$\,\kms, is 0.9\%.  Another
polarization peak at 1\% is observed at $4441\pm15\AA$, without an
obvious counterpart in the flux spectrum.  It may correspond to the
$4584\AA$\ line, in the same multiplet, at the velocity determined
above.

\subsubsection{\ion{O}{1}}
\label{sec:analysis:intrinsic:oi}
A weak polarization signature is associated with \ion{O}{1} $\lambda
7774$\, measured to be $0.50\pm0.13\%$ at $17\pm18\degr$ and $0.3\pm
0.13\%$ at $155\pm4\degr$, at the first and second epochs
respectively.  The large errors are due to the coincident telluric
feature, but still imply the polarization of \ion{O}{1} is not as large as
that observed for \ion{He}{1} and \ion{Ca}{2}.

\subsubsection{\ion{Si}{2}}
\label{sec:analysis:intrinsic:siii}
At the first epoch the \ion{Si}{2} feature is polarized, although with
a peculiar polarization signature tracing the shallow line profile in
the flux spectrum.  At the absorption minimum ($-7\,790$\kms) the
corresponding polarization is $0.50\pm0.15\%$.  The maximum in
polarization ($0.73\pm0.13\%$) actually occurs significantly redward of the
minimum at 6260\AA\,($-4\,490$\kms).  Following \citet{2008arXiv0805.1188M} and \citet{2008arXiv0805.2201M}
(\S\ref{sec:results:spec}), this feature may be a blend of \ion{Si}{2}
and a HV component of $\mathrm{H\alpha}$.  If the polarization maximum is due to $\mathrm{H\alpha}$, rather than \ion{Si}{2}, it would imply that the absorption minimum of the $\mathrm{H\alpha}$ HV component has a velocity of $-17\,050$\kms.  A similar shallow
absorption profile was observed for SN~2005bf, although significant
polarization was detected at the absorption minimum corresponding to
$H\alpha$
rather than \ion{Si}{2}\ \citep{maund05bf}.\\
As discussed in \S\ref{sec:results:spec}, the \ion{Si}{2} feature is
no longer apparent in the flux spectrum of the second epoch.  In
addition, we note that there is no significant signature in the
polarization spectrum in the wavelength range where \ion{Si}{2} is
expected to lie, such that the feature has not been blended with
another feature and instead it has disappeared at that
time.\\
\subsection{Decomposition of the Polarization}
\label{sec:analysis:decomp}
\begin{figure}
\includegraphics[width=4.5cm, angle=-90]{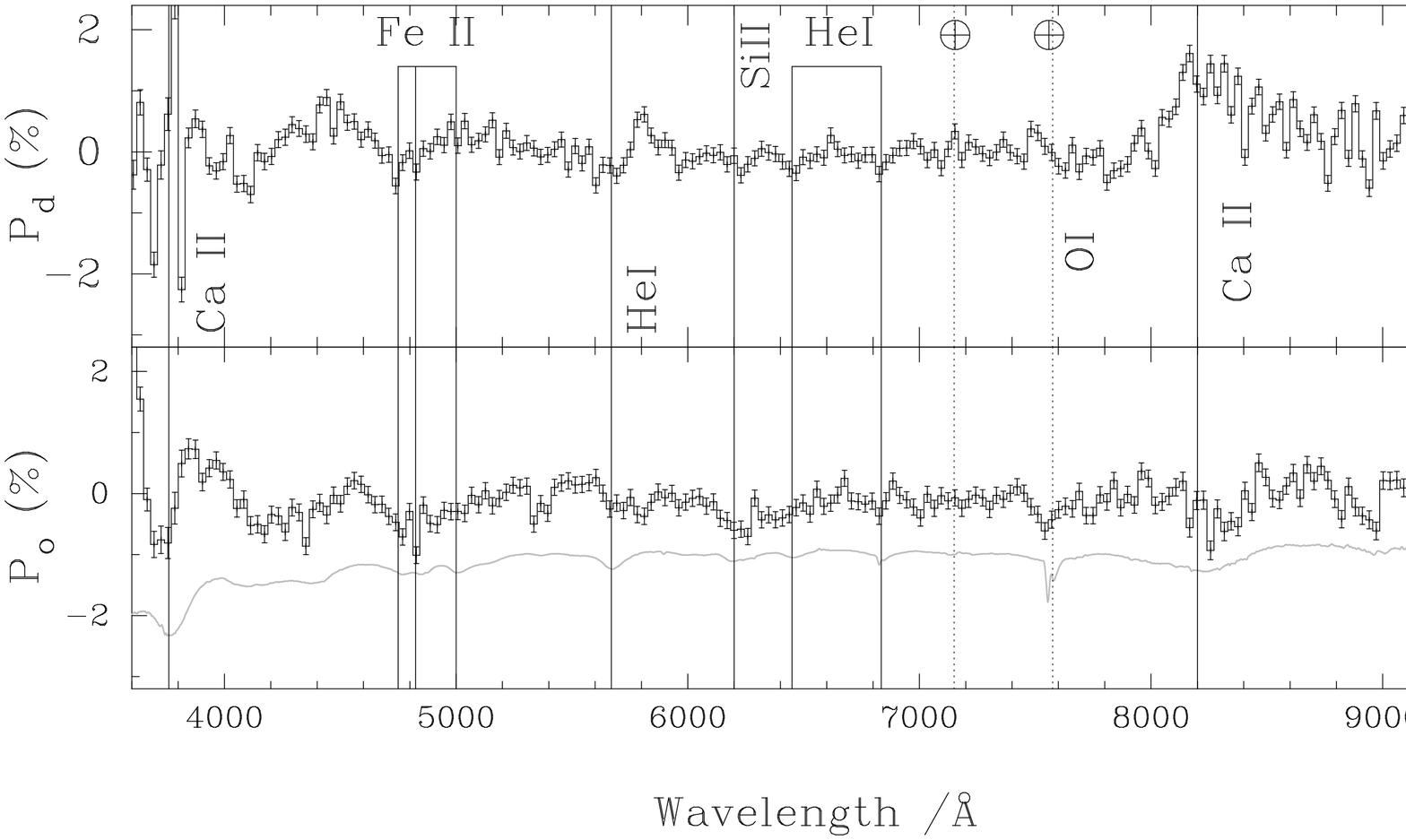}
\caption{Spectropolarimetry of SN~2008D at 2008 Jan 31.22, projected onto the dominant ($P_{d}$; top panel) and orthogonal ($P_o$; bottom panel) axes.  For comparison, the flux spectrum is plotted in grey in the bottom panel.  Line identifications are those from Fig. \ref{fig:obs:epoch1}.}
\label{fig:analysis:epoch1}
\end{figure}

\begin{figure}
\includegraphics[width=4.5cm, angle=-90]{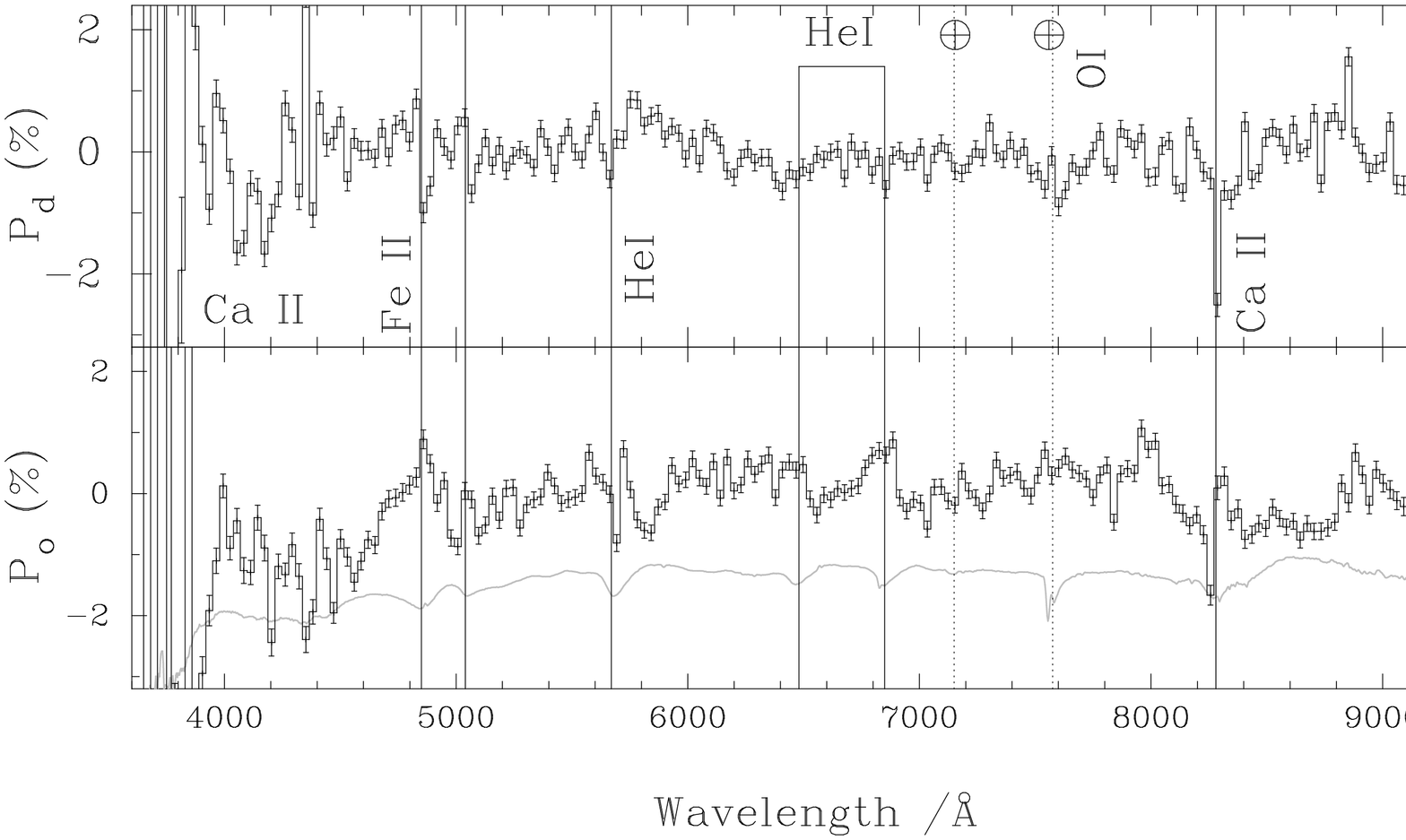}
\caption{Spectropolarimetry of SN~2008D at 2008 Feb 15.18, projected onto the dominant ($P_{d}$; top panel) and orthogonal ($P_o$; bottom panel) axes.  For comparison, the flux spectrum is plotted in grey in the bottom panel.  Line identifications are those from Fig. \ref{fig:obs:epoch2}.}
\label{fig:analysis:epoch2}
\end{figure}

 Following the scheme of \citet{2001ApJ...550.1030W}, the dominant axis of the data on the Stokes $Q-U$ plane was determined by the weighted least-squares fit of a straight line to the data at both epochs.  The angle $\alpha$, between the polarization angle of the dominant axis and the Stokes $Q$ axis were determined to be $40.2\pm1.4\degr$ and $77.3\pm1.3\degr$ at the first and second epochs, respectively.  The dominant ($P_{d}$) and orthogonal ($P_{o}$) Stokes parameters were determined by rotating the data about the origin of the Stokes $Q-U$ plane by an angle $-\alpha$, where:  
\begin{equation}
\left(\begin{array}{c}
P_{d} \\
P_{o}
\end{array}
\right)=
\left( \begin{array}{cc}
\cos 2\alpha & \sin 2\alpha \\
-\sin 2\alpha & \cos 2\alpha \\
\end{array} \right)
\left( \begin{array}{c}
Q-Q_{ISP} \\
U-U_{ISP} \end{array} \right) .
\end{equation}
The determination of the polarization along and perpendicular to the dominant axis provides information on the principal asymmetry of the SN, and deviations from it (e.g. by the line forming regions of particular species in the spectrum) and removes the effects of the position angle at which the SN was observed (although certain inclination effects remain).  The spectropolarimetry projected into the dominant and orthogonal Stokes parameters is shown as Figs. \ref{fig:analysis:epoch1} and \ref{fig:analysis:epoch2}.  In the case of SN~2008D, like SN~2002ap \citep{2003ApJ...592..457W}, at both epochs the dominant axes were found to not pass through the origin of the Stokes $Q-U$ plane, such that the presence of continuum polarization is observed as an offset in the orthogonal polarization component.\\
At the first epoch, \ion{Si}{2}/$\mathrm{H\alpha}$ and \ion{He}{1} lines are predominantly polarized along the dominant axis, whereas \ion{Ca}{2} H\&K and IR3 and the \ion{Fe}{2} have components in both the dominant and orthogonal directions. A strong orthogonal polarization is observed close to the expected position of the \ion{O}{1} $\lambda 7774$ absorption.\\
At the second epoch the situation is reversed, with the \ion{He}{1} lines being stronger in the orthogonal component.  Despite the apparent decrease in the strength of the \ion{Fe}{2} absorptions, these features are still present in the dominant component, although the strong feature observed in $P_{o}$ is due to \ion{He}{1} $\lambda 5015$.\\
Caution is encouraged in interpreting the data on the dominant and orthogonal axes.  The apparent shift of \ion{He}{1} from the dominant axis (first epoch) to the orthogonal axis (second epoch) is primarily due to the apparent change in angle of the dominant axes ($\Delta \alpha \sim 37\degr$) between the two epochs, due to the evolving flux and polarization of lines in the spectrum between the two epoch; as opposed to a bulk change in the polarization properties of the SN between the two epochs (such as observed for SN~2001ig; \citealt{my2001ig}).   Examination of the data on the Stokes $Q-U$ plane in Figs. \ref{fig:obs:epoch1qu} and \ref{fig:obs:epoch2qu} shows the data to be clustered in the same location at the two epochs.\\
As discussed in \S\ref{sec:analysis:geometry}, the relative polarization of different species in the dominant and orthogonal directions indicates that, on the sky, the line forming elements have very different distributions relative to eachother within the ejecta.
\subsection{The polarization angles of absorption features}
\label{sec:analysis:geometry}

\begin{figure}
\includegraphics[width=9cm, angle=0]{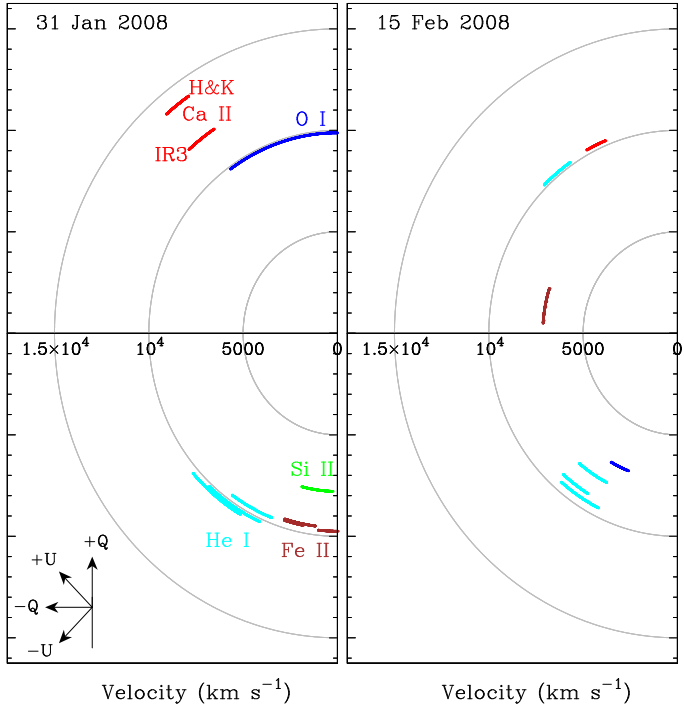}
\caption{Polar plots showing the velocity and polarization angle at absorption minimum for \ion{Ca}{2} H\&K, \ion{He}{1} $\lambda\lambda 5015,5876,6678,7065$, \ion{Fe}{2} (42), \ion{Si}{2} $\lambda 6355$, \ion{O}{1} $\lambda 7774$ and \ion{Ca}{2} IR3 for the first and second observations of SN~2008D.  The angle of each arc corresponds to the observed polarization angle on the sky (the arc length is $2\Delta\theta$) and the radius of the arc corresponds to the velocity.  In this figure, North is up and east is left, with the corresponding orientation of the Stokes parameters indicated in the bottom left of the left panel.}
\label{fig:analysis:polar}
\end{figure}

The true power of spectropolarimetry, to study the shapes of SNe, is its ability to provide information of the geometry in the plane of the sky.  Comparison of the degree of polarization and polarization angle against the velocities observed in the flux spectrum permits a three dimensional tomographic analysis indicating the distribution of various elements within the ejecta in addition to the shape of the photosphere.\\
The polarization angles and velocities at absorption minimum for important lines in the spectrum are presented for both epochs in Fig. \ref{fig:analysis:polar}.  The polarization angle measured at the absorption minimum corresponds to the polarized component of continuum light that is {\it not} absorbed by the line forming region, hence the approximate distributions of the various line forming regions on the sky are rotated $\pi / 2$ to the positions shown in Fig. \ref{fig:analysis:polar}.\\
The observed polarization angle will be the result of mixing of polarized components from the continuum (if the photosphere is asymmetric) and the line's effective polarization (due to the shadow it casts on the photosphere).  In the case of SN~2008D we have established, however, that the observed line polarization is significantly larger than the continuum polarization, and will dominate the measured polarization angle.\\
As shown in Fig. \ref{fig:analysis:polar}, at the first epoch \ion{He}{1} and \ion{Fe}{2} are found in similar parts of the ejecta, with similar velocities ($\sim -10\,000$\kms), while the \ion{Si}{2} $\lambda 6355$ line is formed in a similar location, but at lower velocities.  Both the \ion{Ca}{2} H\&K and IR3 lines have the same polarization angle, but are offset from the \ion{He}{1} lines by $\Delta \theta = -100\degr$ (or $+80\degr$).  This implies that there are, at least, two distinct ejecta components and that the \ion{Ca}{2} lines, at much higher velocities, are disjoint from the rest of the ejecta.\\
At the second epoch,  all velocities have decreased and \ion{Ca}{2} and most of the \ion{He}{1} lines are
observed at approximately the same position angles as at the first epoch.
The feature at $\lambda 5876$ is separate from the other \ion{He}{1}
lines, with a polarization angle more consistent with \ion{Ca}{2}.  At
this epoch, the \ion{He}{1} $\lambda 5015$ line is visible, and is
grouped with the higher wavelength lines at $\lambda 6678,7065$.  This
suggests that the feature associated with $\lambda 5876$ is actually a
blend with \ion{Na}{1}D \citep{2003ApJ...592..457W}.  \ion{Fe}{2} (42)
is observed at an intermediate
polarization angle.\\
It is unclear at the first epoch, given the large degree of
uncertainty, whether the \ion{O}{1} line forming region occupies a
similar space to the \ion{Fe}{2} line forming region or if it is
consistent with the \ion{Ca}{2} region (due to the fact that a full
rotation of the Stokes parameters occurs over only $180\degr$).  There
is significant uncertainty in both the degree of polarization and
polarization angle associated with \ion{O}{1} due to coincidence with
a strong telluric feature.  As polarimetry measures differences in
flux, the telluric absorption (assumed to be unpolarized) should not
have any effect on the measured polarization, except for a reduction
in the signal-to-noise ratio.
\section{Discussion}
\label{sec:discussion}
\subsection{Spectropolarimetric Classification of SN~2008D}
\label{sec:discussion:class}
Following the classification scheme of \citet{2008ARA&A..46..433W}, SN~2008D/XRF~080109 is classified as being of type D1+L(\ion{He}{1})+L(\ion{Ca}{2}).  This classification reflects the fact that a dominant axis can be fit to the data, although the polarization is better described as an ellipse on the Stokes $Q-U$ plane.  Furthermore, distinct loop features have been identified for \ion{He}{1} and \ion{Ca}{2}.
\subsection{Comparison with other Type Ibc SNe}
\label{sec:discussion:typeibc}
\begin{figure}
\includegraphics[width=9cm,angle=-90]{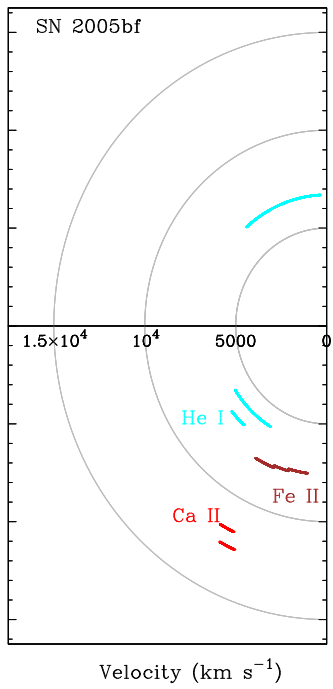}
\caption{Polar plot of the polarization angles of particular spectral features from the observation of SN~2005bf at 6 days prior to second maximum \citep{maund05bf}.}
\label{fig:discussion:05bf}
\end{figure}

SN~2008D shows a number of similarities and differences with the
spectropolarimetry of other CCSNe, which can be used to compare and
contrast likely geometries of these events.  The intrinsic level of
polarization of the continuum, dependent on the shape of the
photosphere, is significantly detected but low, suggesting asymmetries
of $\sim 10\%$.  Similar low levels of continuum polarization have
been observed over the range of Type Ibc SNe, including 2002ap,
2005bf, 2006aj and 2007gr
\citep{2002PASP..114.1333L,2002ApJ...580L..39K,2003ApJ...592..457W,maund05bf,2007A&A...475L...1M,2008arXiv0806.1589T}.
The starkest contrast between these events is, however, the
polarization associated with specific spectral lines.\\
For SN~2002ap, significant polarization associated with the \ion{O}{1}
$\lambda {7774}$ (0.9\%; \citealt{2003ApJ...592..457W}) and
\ion{Ca}{2} IR3 \citep{2002ApJ...580L..39K,2002PASP..114.1333L}
absorption components was observed. In that case, however, the
progenitor star was a non-degenerate C/O star without He, whereas for
SN~2008D the presence of a strong \ion{He}{1} feature specifically
implies the explosion of He-rich star
\citep{2008Natur.453..469S,2008arXiv0805.2201M}.  \citet{maund05bf} interpreted the lack of polarization associated with \ion{O}{1} in spectropolarimetry of the Type Ib SN~2005bf as indicating ``shielding'' of the O-rich core by the He envelope.  On the other hand, 2007gr was a Type Ic SN, for which no He layer expected, which also showed little polarization associated with \ion{O}{1} \citep{2008arXiv0806.1589T}.\\
\citet{2008arXiv0805.2201M} argue against using SN~2005bf as the
comparison for SN~2008D on photometric and spectroscopic grounds, due
to the double peaked light curve and the appearance of \ion{He}{1} in
the spectrum at 6 days prior to the second light curve maximum ($\sim
34$ days post-explosion).  We note that the polarization properties of
SN~2008D and SN~2005bf are actually quite similar. The loop feature
for the \ion{He}{1} $\lambda 5876$ line for SN~2008D at the second
epoch is similar to that observed for SN~2005bf \citep{maund05bf};
except in this instance the signal-to-noise ratio is lower, making the
loop less clear in Fig. \ref{fig:obs:heline}.  Further similarity is
observed for the resolved lines of \ion{Fe}{2} in the flux spectrum,
the low level of \ion{O}{1} line polarization and the presence of
\ion{Si}{2} and HV $\mathrm{H\alpha}$\ absorption features.
Importantly, \ion{Ca}{2} lines of both SN~2008D and 2005bf are
observed to have minima at substantially higher velocities than
\ion{He}{1}, \ion{O}{1} and \ion{Fe}{2} (see
Fig. \ref{fig:discussion:05bf}) and are significantly polarized.  This
suggests that, at least for the second light curve maximum, SN~2005bf
is similar to SN~2008D
at light curve maximum for the purposes of comparing their polarimetry and, ultimately, the natures of their explosions.\\
HV Ca components have proven quite common in Type Ia SNe (SN 2001el -
\citealt{2003ApJ...591.1110W}; SN~2006X - \citealt{nando2006X};
\citealt{2005ApJ...623L..37M}); here we note that they also seem
rather common in SNIb/c as well.  Spectropolarimetry has shown,
however, that these features are highly polarized and that the
polarization angles of these features are substantially different from
the rest of ejecta.
\subsection{Inferred geometry of the ejecta}
\label{sec:discussion:geometry}
Our spectropolarimetry of SN~2008D shows clear evidence that the ejecta are {\it chemically segregated}.\\
Fig. \ref{fig:analysis:polar} shows that the \ion{Ca}{2}
H\&K and IR3 lines are formed at higher velocities than all the other
spectral features studied in the spectrum.  The \ion{He}{1} lines all have
approximately similar polarization angles at both epochs, except at
the second epoch where one feature ($\lambda 5876$) is observed to
have a polarization close to that of
\ion{Ca}{2}.\\
The observed maximum polarizations for \ion{He}{1} and \ion{Ca}{2},
which exceed the theoretical limit for simple blocking of the
unpolarized light, implies that the line forming regions for these
species are not distributed evenly across the photosphere.  The HV
components of the \ion{Ca}{2} lines at both epochs, along with the
significantly different polarization angle to the rest of the
constituents of the ejecta, suggest the \ion{Ca}{2} may be formed in
a clump with a significant projected velocity along the line-of-sight.
For SN~2007gr, \citeauthor{2008arXiv0806.1589T} inferred that the HV
calcium was constrained in a bipolar distribution, with the other
elements such as \ion{O}{1} forming a torus in the orthogonal plane
(due to the different polarization angles).  Such a bipolar
distribution of \ion{Ca}{2}, for these Type Ibc SNe, may be
related to the action of a jet-like flow.  The low continuum polarization may indicate that the ejecta are viewed along, or close to, a principal axis of symmetry for the ejecta.\\
The presence of loop structures has been observed for nearly all CCSNe
\citep{1988MNRAS.231..695C,2003ApJ...591.1110W,maund05bf} and some
Type Ia SNe.  These features have been interpreted by
\citet{2003ApJ...593..788K} and \citet{maund05bf} as being due to the
line-forming region, of the specific species, having an axial symmetry
inclined with respect to the axial symmetry of the photosphere.  This
implies that there is no single global axial symmetry, and rather that
the geometries of the ejecta are different for clumps of different
species.  If a single axial symmetry were present, then the data would
be well described by a single dominant axis, and for SN~2008D it has
been shown (see \S\ref{sec:analysis:decomp} and
Figs. \ref{fig:analysis:epoch1} and
\ref{fig:analysis:epoch2}) to not be the case.\\
The low level of continuum polarization determined here suggests the
deviation of the photosphere from a spherical symmetry is low.  This
implies that the photosphere is receding through the ejecta at an
approximately isotropic rate, which further suggests that the
photosphere did not cross a significant change in the excitation
structure between the two epochs of spectropolarimetry.  In the cases
of SNe 2004dj \citep{2006Natur.440..505L}, SN~2001ig \citep{my2001ig}
and 2001dh \citep{my2001dh} significant changes in the degree of
polarization and the polarization angle with time implied depth
dependent geometry discontinuities.  In the case of Type IIP and Type
IIb SNe, such a significant change in polarization is expected once
the photosphere crosses the chemical boundary between the H and He layers, where the
photosphere can recede more quickly since \ion{He}{1} is less ionized
and, hence, presents less opacity than H.  The geometry of \ion{He}{1}
is expected, therefore, to more closely match the underlying Ni
distribution due to the asymmetric apparent luminosity of the
photosphere \citep{2001AIPC..586..459H}.  Another such discontinuity
is expected when the core products of nucleosynthesis and explosive
burning are revealed, for which hydrodynamic simulations predict
complex
three-dimensional structures \citep{2008arXiv0812.3918C}.\\
\citet{2008arXiv0810.4333G} observed small changes in the broad-band
polarization properties of SN~2008D over a larger time frame, from
2008 Jan 13 to 2008 Mar 28, which are consistent with our synthetic
V-band polarimetry (see \S\ref{sec:results:pol}).  While the true
stability of the polarimetric standard adopted by
\citeauthor{2008arXiv0810.4333G}, itself a SN, is debatable, the  uncorrected polarimetry of SN~2008D shows only small variability. This implies that a significant geometry
discontinuity was not just unobserved due to the paucity of our own
observations.  Our spectropolarimetry suggests, however, that while
some of the variability may truly arise in the continuum, an
additional factor that may contribute to the measured broad-band
polarization is evolving line polarization.  The most important
feature possibly affecting the V-band polarization is the
\ion{He}{1}/\ion{Na}{1} feature, which is stronger, narrower and more
polarized at the second epoch compared to the first.  The striking absence of a discrete
geometry discontinuity suggests that any axisymmetric structure
directly associated with the explosion mechanism is not observed at
the times of our observations or those of \citet{2008arXiv0810.4333G}, and rather that the continuum
polarization arises from only an elongation of the core, similar to
that observed at early times for
SN~2002ap \citep{2003ApJ...592..457W}.\\

\subsection{SN~2008D in the jet-torus paradigm}
\label{sec:discussion:jettorus}
\begin{figure}
\includegraphics[width=9.5cm]{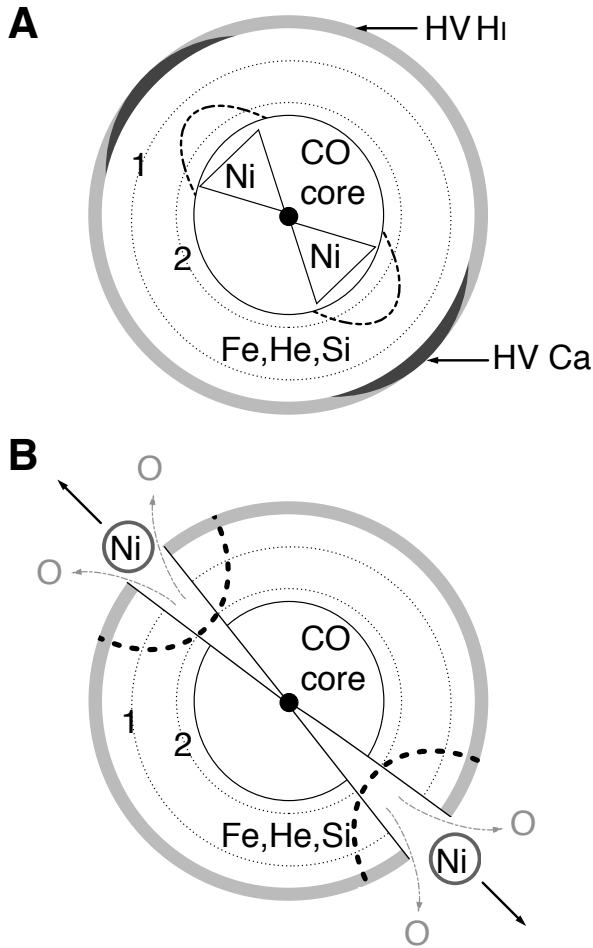}
\caption{Possible scenarios, in the jet-torus paradigm, for the
  geometry of SN~2008D. The location of the HV line forming regions
  are indicated by the light grey zone (and dark grey zones for
  \ion{Ca}{2}).  A) Thermal energy dominated jet explosion.  The jet
  elongates the core (heavy dashed line).  HV \ion{Ca}{2} and
  $\mathrm{H\alpha}$ features arise in a shell at the edge of the
  ejecta, with the high velocity material aligned with the jet axis.
  \ion{He}{1}, \ion{Si}{2} and \ion{Fe}{2} occur in the main portion
  of the ejecta.  The Ni-rich jets are contained within core, below
  the depth of the photosphere at the first and second epochs (dotted
  line).  B) Kinetic energy dominated jet explosion.  Ni-rich material
  and other core-material, such as \ion{Ca}{2} and \ion{O}{1}, are
  transported to the surface.  The outer edges of the ejecta are
  excited in the vicinity of the Ni-rich clumps (heavy black dotted
  lines).  The position of the photosphere, at the two observational
  epochs, are indicated by light dotted lines.}
\label{fig:discussion:scenario}
\end{figure}

The apparent geometry of SN~2008D, at the two epochs, suggests the
possibility of a jet induced explosion.  In the jet-torus explosion
model, a jet-like flow propagates through the progenitor.  Such a flow
may originate from an asymmetry of neutrino deposition in the ejecta,
the Standing Accretion Shock Instability \citep{2007Natur.445...58B}
or a magneto-rotational instability \citep{2008ApJ...683..357M}. HV
material is expected to arise from material impacted by the jet, while
lateral shocks compress the main portion of the ejecta into a toroidal
configuration \citep{1999ApJ...524L.107K}.\\
The lack of a detected geometry discontinuity is suggestive that
either a jet-like flow had already been revealed at the time of our
first observation or that it still remained shielded inside the core
at the time of our second observation.  The properties of jets which
can either punch through the ejecta or stall within the core have been
studied previously, and shown to be dependent on the relative mix of
thermal and kinetic energy in the jet
\citep{2001AIPC..586..459H,2008arXiv0812.3918C}.  We propose two models (Scenarios A and B; schematic diagrams of these configurations are presented as Fig. \ref{fig:discussion:scenario}), in the jet-torus paradigm, which may have given rise to the spectropolarimetric properties:\newline
{\it Scenario A:} The jet-like flow (thermal energy dominated) stalled
within the core.  High-velocity \ion{Ca}{2} is produced approximately
along the direction of the jet-like flow, while the main portion of
the ejecta exterior to the photosphere at both epochs (\ion{He}{1},
\ion{Fe}{2}, \ion{O}{1} and \ion{Si}{2}) has a squashed
spherical/toroidal geometry orthogonal to that of \ion{Ca}{2}.
\citet{maund05bf} constructed a tilted jet model to explain the
apparent lack of a single global axial symmetry for SN~2005bf.
Similar to SN~2002ap \citep{2003ApJ...592..457W}, the low degree of
asymmetry of the photosphere is taken as an indication that such jets
had not reached the height defined by the photosphere as measured at
the epochs of the two observations.  On the other hand the significant
polarization associated with particular spectral features, with
differing polarization angles for different species, is indicative of
differing distributions of these elements within the ejecta, such that
a single spherical symmetry is not appropriate.  That loops are also
observed for some spectral features is perhaps supportive of SN~2008D
having a similar configuration as SN~2005bf, and a tilted axis of
excitation with respect to the axis of symmetry of the photosphere.
This may be evidence that there has not been full scale transport of
material that would have led to homogeneity, and that the elements
reside in different portions of the ejecta.  This scenario can be
directly tested with observations at later epochs, when a large
increase in the continuum polarization is expected as the asymmetric
core-layers, containing jet material, are revealed.
\\
{\it Scenario B:} The jet-like flow (kinetic energy dominated)
completely broke out of the progenitor prior to our first
spectropolarimetric observation (and those of
\citeauthor{2008arXiv0810.4333G}, at 18 days prior to our first
observation), such that no discontinuity would be apparent as in the
case of the jet stalling within the core.  Such a scenario was
suggested for SN~2002ap, where a polarization peak in the vicinity of
\ion{O}{1}$\lambda 7774$/\ion{Ca}{2} IR3 was interpreted as being due
to light reflected by electrons in a high-velocity Ni-rich jet-like
clump, exterior to the main ejecta, moving at $\gtrsim 0.115c$
\citep{2002PASP..114.1333L,2002ApJ...580L..39K}.
\citet{2003ApJ...592..457W} rejected this picture because: 1) the
polarization was associated with \ion{O}{1} moving at $\sim 0.07c$ and
2) no features due to Ni or Co (in the range 1.6-1.8 $\mu m$) were
present in the flux spectrum at early times, as would be expected if
these radioactive elements had been deposited outside the main ejecta.
For SN~2008D we note similarly that \citet{2008arXiv0805.2201M} did
not detect Ni or Co features in their IR spectra nor do we detect the
asymmetric excitation of the ejecta illuminated by exterior Ni-rich
clumps (with an obviously bipolar distribution). Similarly, the low
degree of \ion{O}{1} polarization, suggests there was little/no
transport of core-material to the outer layers of the ejecta by the
action of a jet-like flow.\\
We conclude that a jet-like flow, if present, stalled inside the CO core and, therefore, support Scenario A.  An important consequence of Scenario A is, however, that the HV \ion{Ca}{2} is {\it not} formed from products of nucleosynthesis from the progenitor, that was transported to the high-velocity (large radius) layer by a jet-like flow, rather it arose from primordial Ca abundance of the progenitor.\\
The question remains, however, as to the nature of the control
mechanism which dictates why some Type Ibc SNe are associated with the
GRB phenomenon and, hence, with highly-energetic jets which penetrate
to the exterior of the progenitor and others, such as SNe 2008D and
2005bf show evidence for jets which may have only remained in the core
\citep{2001AIPC..586..459H}.

\subsection{Implications for the shock-breakout scenario}
\label{sec:discussion:shock}
The suggestion from spectropolarimetric observations that any jet-like
flows did not completely breakout has important implications for the
interpretation of the X-ray emission of the associated XRF 080109.\\
In addition to our spectropolarimetry, \citet{2008arXiv0805.2201M} report peculiar emission line profiles in
late-time (91d after max.) nebular spectra of SN~2008D, at an epoch
when the SN is optically thin, which is claimed to trace the asymmetric ejecta
\citep[see also][]{2008Sci...319.1220M}.  We note that our
early time spectropolarimetric observations of asymmetries in the
distributions of a number of species within the ejecta are consistent
with this picture (for example \ion{O}{1}).  At such late times,
however, the ejecta can be shaped by a number of competing factors:
the interaction between the Circumstellar Medium (CSM) and the ejecta
and the evolution and interaction of physical processes (such as
shocks and instabilities) over the lifetime of the SN.  Importantly,
at early times the measured geometry reflects the asymmetric nature of
the explosion mechanism with little evolution from the moment of
explosion.  In this case, the observed geometry is more closely
related to the geometry from which
the early X-ray emission originated than the late-time observations of \citet{2008arXiv0805.2201M} may be.\\
There are a number of open issues regarding the early X-ray emission
from SN 2008D: whether it is due to thermal emission of shock break
out \citep{2008arXiv0806.0371C} or not
\citep{2008Natur.453..469S,2008arXiv0801.4325X,2008MNRAS.tmp..741L};
whether the emission is non-thermal
(\citeauthor{2008Natur.453..469S}), a single black body
(\citeauthor{2008arXiv0806.0371C}), or multiple black bodies
(\citeauthor{2008MNRAS.tmp..741L}); whether there is a dense wind
surrounding the progenitor (\citeauthor{2008Natur.453..469S}) or not
(\citeauthor{2008arXiv0806.0371C}); whether non-thermal emission
arises from a shock-mediated Fermi acceleration
(\citeauthor{2008Natur.453..469S}) or a variation of the
forward/reverse shock paradigm of GRBs
(\citeauthor{2008arXiv0801.4325X}). The common feature of these
analyses is that they assume, or conclude, that the associated
processes are spherically symmetric, or nearly so.  Our
spectropolarimetry data show that this is not the case, certainly by
nearly maximum light when the observations are made, and presumably
not as the shock is erupting from the surface of the star and
propagating into any circumstellar medium.\\
Our observations do not directly constrain the geometry at breakout,
but they give a strong caution that asymmetric shock breakout should
be considered. Asymmetric shock breakout is unlikely to be accompanied
by a single black body spectrum, so arguments that a single black body
does not fit the X-ray spectrum are questionable. We also note that
the X-ray spectrum softens considerably over the course of the X-ray
outburst (\citeauthor{2008Natur.453..469S}, supplemental material), so
using the mean X-ray spectrum to characterize the physical nature of
the outburst requires caution. Arguments that the X-ray burst cannot
be a shock break out
(\citeauthor{2008arXiv0801.4325X,2008MNRAS.tmp..741L}) are subject to
uncertainties in the color temperature compared to the effective
temperature (\citeauthor{2008arXiv0806.0371C}), so must also be
considered with caution. Given the concrete evidence for asymmetry and
the likelihood that the shock erupted from restricted areas of the
surface of the progenitor, all of these issues need to be reconsidered
in the context of jet-like shock breakout.

\section{Conclusions}
\label{sec:conclusions}

SN~2008D has been observed to be significantly polarized at maximum
and two-weeks after maximum light.  A significant amount of
polarization is associated with the Interstellar Polarization,
$p_{ISP}=1.2\%$ and $\theta_{ISP}=135 \pm 4.3 \degr$, arising in the
host galaxy.\\
SN~2008D is intrinsically polarized, with a continuum polarization
implying an asymmetry of the photosphere of $\lesssim 10\%$.  Strong
polarization associated with spectral lines shows evidence for
significant asymmetries in the line forming region of the ejecta above
the photosphere.  Loops on the Stokes $Q-U$ plane are observed for
lines of \ion{He}{1}, in particular $\lambda 5876$, and \ion{Ca}{2}.
A low level of polarization is potentially associated with \ion{O}{1}
$\lambda 7774$, although this feature is complicated by blending with
a strong telluric feature.  Significant polarization is also observed
for \ion{Fe}{2} and \ion{Ca}{2}; in the case of the latter,
polarization of the highest velocity component of IR3 was measured to
be $1.8\pm0.3\%$ and $2.5\pm 0.7\%$ at the first and second epochs
respectively.  The polarization of the \ion{Ca}{2} IR3 feature shows a
significantly different polarization angle, at both epochs, to other
species in the spectrum, suggesting it arises in a different spatially
distinct, higher-velocity portion of the ejecta than the
lines of other species (i.e. the line forming region for \ion{Ca}{2} does not just lie at higher velocities than the line forming regions of other species, but also with a different geometry).\\
The strong polarization associated with these lines suggests
incomplete coverage of the photosphere by the line forming regions for
these species, and the differing polarization angles imply that the
line forming regions are different for different species.  The absence
of a strong polarization signature associated with \ion{O}{1} $\lambda
7774$, the absence of Ni or Co in early time IR spectra, and a
relatively low level of continuum polarization suggests no core
material was
transported to the outer layers of the ejecta by, for example, the action of a jet-like flow.\\
The observed asphericity of the photosphere (albeit at a relatively low level) and the chemically dependent structure of the line forming regions above the photopshere demonstrates that SN~2008D was not spherically symmetric.\\
If SN~2008D was the result of a jet-induced explosion, then we
conclude that the jet stalled within the core, causing the core and,
hence, photosphere to become slightly elongated.  The degree of
elongation is observed to change as a function of depth over the
period of our spectropolarimetry and over the period for which
broad-band polarimetry has also been reported.  In terms of
spectropolarimetry, SN~2008D is found to be closely related to
SN~2005bf (albeit at 6 days before the second light curve maximum of
the latter).  The spectropolarimetric study of the geometries of
SN~2002ap, 2005bf, and 2008D are shown to be consistent within the
jet-torus paradigm, but in which the jet
remains within the cores of the progenitor.  We hypothesise that relatively late-time polarimetric observations of CCSNe, before the optical depth to electron scattering drops below unity, will reveal a geometry discontinuity consistent with the photosphere having reached the position of a stalled jet.\\
The observation of asymmetries of SN~2008D at early times has
important implications for the study of the X-ray and UV shock
breakout and the assumption of spherical symmetry.

\section*{Acknowledgements}
The research of JRM is funded through the Sophie \& Tycho Brahe
Fellowship.  The Dark Cosmology Centre is supported by the DNRF.  The
research of JCW is supported in part by NSF grant AST-0707769.  The
authors are grateful to the European Organisation for Astronomical
Research in the Southern Hemisphere for the generous allocation of
observing time. They especially thank the staff of the Paranal
Observatory for their competent and never-tiring support of this
project in service mode.

\bibliographystyle{apj}
%\bibliography{/Users/justyn/Documents/Bibliography/main}
%%%%%%%%%%%%%%%%%%%%%%%%%%%%%%%%%%%%%
%Tables
%%%%%%%%%%%%%%%%%%%%%%%%%%%%%%%%%%%%%
\begin{table}
\caption{\label{tab:obs:log}Journal of Spectropolarimetric Observations of SN~2008D/XRF 080109.}
\begin{tabular}{cccccc}
\hline\hline
Object   &  Date &  Exposure &  Median  &  Median   & Type  \\
         &  UT   &   (s)     &  Airmass &  Altitude &       \\
\hline
LTT~3218 &2008 Jan 31.16 &   30         & 1.027 & 76.74& Flux Std.\\
SN~2008D &2008 Jan 31.22 & $8\times750$ & 1.894 &31.80 & Object     \\
\\
SN~2008D &2008 Feb 15.18 & $8\times900$ & 1.899 &31.69 & Object      \\
GD~108   &2008 Feb 15.23 & 50           & 1.056 &71.33 &Flux Std.\\
Vela~1~95&2008 Feb 15.24 & $4\times30$  & 1.156 &59.93 &Pol. Std.\\ 
\hline\hline
\end{tabular}
\end{table}

%%%%%%%%%%%%%%%%%%%%%%%%%%%%%%%%%%%%%
%Figures
%%%%%%%%%%%%%%%%%%%%%%%%%%%%%%%%%%%%%





\end{document}